\documentclass{aa}  
\usepackage{graphicx}
\usepackage{txfonts}
\usepackage{natbib}
\usepackage{multirow}
\bibpunct{(}{)}{;}{a}{}{,} % A&A style

\begin{document}

\title{Calibration of the pre-main sequence RS Cha binary system}
%   \subtitle{}

\author{
E. Alecian 
\inst{1}\and
M-J. Goupil 
\inst{1}\and
Y. Lebreton 
\inst{2}\and
M-A. Dupret
\inst{1}\and
C. Catala 
\inst{1}
}

\offprints{E.Alecian(evelyne.alecian@obspm.fr)}

\institute{
%1
Observatoire de Paris, LESIA, 5, place Jules Janssen, F-92195
Meudon Principal CEDEX, France \email{evelyne.alecian@obspm.fr}\and
%2
Observatoire de Paris, GEPI, 5, place Jules Janssen, F-92195
Meudon Principal CEDEX, France
}

   \date{Received ; accepted }

% \abstract{}{}{}{}{} 
% 5 {} token are mandatory
 
  \abstract
  % context heading (optional)
  % {} leave it empty if necessary  
{The calibration of binary systems with accurately known masses and/or radii provides powerful tools to test stellar structure and evolution theory and to determine the age and helium content of stars. We study the eclipsing double-lined spectroscopic binary system RS Cha, for which we have accurate observations of the parameters of both stars (masses, radii, luminosities, effective temperatures and metallicity).}
  % aims heading (mandatory)
   {We have calculated several sets of stellar models for the components of the RS Cha system, with the aim of reproducing simultaneously the available observational constraints and to estimate the age and initial helium abundance of the system.}
  % methods heading (mandatory)
   {Using the CESAM stellar evolution code, we model both components starting from the initial mass and metallicity and adjusting the input parameters and physics in order to satisfy the observational constraints.}
  % results heading (mandatory)
   {We find that the observations cannot be reproduced if we assume that the abundance ratios are solar but they are satisfied if carbon and nitrogen are depleted in the RS Cha system with respect to the Sun. This is in accordance with the abundances observed in other young stars. The RS Cha system is in an evolutionary stage at the end of the PMS phase where models are not strongly sensitive to various physical uncertainties. However we show that the oscillations of these two  stars, which have been detected, would be able to discriminate between different options in the physical description of this evolutionary phase.}
  % conclusions heading (optional), leave it empty if necessary 


   \maketitle

%
%________________________________________________________________

\section{Introduction}

Stellar calibration that-is determining masses, radii, ages of stars using theoretical stellar evolutionary tracks is a powerful tool. However, its results strongly depend upon the validity of the adopted stellar models. The physical description of these models can be validated with a small set of stars whose masses,  radii, luminosities, effective temperatures and metallicities  are accurately  known.  An  eclipsing binary system is therefore an excellent candidate for  such  a test as the orbital information provides masses and radii and the assumed common origin of both components (implying same age and chemical composition) brings out additional severe constraints to the modelling. Validation of  main sequence (MS) stellar evolution models has been quite extensively  performed (e.g. Noels et al. 1991, Morel et al. 2000). On the other hand, only a few validation works have sofar concerned the pre-main sequence (PMS) phase. Palla \& Stahler (2001) tested their PMS theoretical evolutionary tracks using eight binary systems, assuming a solar metallicity in all of them, although this assumption appeared later on not valid in all but one case. Marques et al. (2004)  modelled the binary system EK Cep using a $\chi^2$-minimisation and obtained the most reliable set of theoretical stellar parameters which reproduces the observational one.  These authors  succeeded in  modelling  the PMS secondary component but failed to obtain a  theoretical radius which reproduces the observed one  for the  MS primary component.

The A-type system RS Chamaeleontis (RS Cha) is an eclipsing double-lined spectroscopic binary system. Mamajek et al. (1999) reported X-ray emission from the $\eta$ Cha cluster, which indicates the PMS status of both components of RS Cha. Up to recently, all fundamental parameters of the components of this system were known, except the metallicity. However,  the knowledge of the metallicity is crucial to model the structure and the evolution of these PMS stars (see Sect.4.3.2). The metallicity of  RS Cha system which is used here has been obtained with spectroscopic data collected at the SAAO (South African Astronomical Observatory) (Alecian et al., 2005; hereafter paperI). We took advantage of these data to redetermine the masses and radii of both stars. The resulting physical parameters of both components are detailed in Table \ref{mrt}. 

We are then able to model these two stars and confront the results to observations. This paper presents an advanced study of the modelling of the pre-main sequence RS Cha system. The next section describes the physical input of our standard evolution models. Sect. 3 states the observational constraints that our RS Cha models must reproduce. As our standard models cannot reproduce the whole set of observations, Sect. 4 investigates  the sensitivity of  PMS models 
to uncertainties in the input physics and physical parameters. Sect.5 presents the  final  calibrated models for the RS Cha system which reproduces all the available observations. Sect.6 contains an analysis of the recently discovered pulsations in both components of RS Cha and discusses  theoretical oscillation modes and periods. Finally conclusions  are presented in Sect. 7.

\begin{table}[t]
\caption{Fundamental parameters of RS Cha. R00: \citet{ribas00}, CN80: \citet{clausen80}, M00: \citet{mamajek00}. P stands for primary and S for secondary}
\label{mrt}
\renewcommand{\multirowsetup}{\centering}
\centering
\begin{tabular}{@{}p{1.9cm} p{1.7cm} p{1.7cm} p{2.1cm}@{}}
\hline\hline
                       &        P        &       S           & References \\
\hline
$M/M_{\odot}$          & $1.89\pm 0.01$  & $1.87\pm0.01$     & paperI \\
$R/R_{\odot}$          & $2.15\pm0.06$   & $2.36\pm0.06$     & paperI \\
$T_{\rm eff}$ (K)      & $7638\pm76$     & $7228\pm72$       & R00 \\
$\log (L/L_{\odot})$   & $1.15\pm 0.09$  & $1.13\pm0.09$     & $L=4\pi R^2\sigma T_{\rm eff}^4$ \\
$\log(g)$(cm.s$^{-2})$ & $4.05\pm0.06$   & $3.96\pm0.06$     & $g=MG/R^2$ \\
$v\sin i$ (km.s$^{-1}$)& $64\pm6$				 & $70\pm6$					 & paperI \\
P (day)                & \multicolumn{2}{c}{1.67}            & paperI \\
i ($^{\circ}$)         & \multicolumn{2}{c}{$83.4\pm0.3$}    & CN80 \\
$\rm[Fe/H]	$	         & \multicolumn{2}{c}{$0.17\pm0.01$}   & paperI \\
\hline
\end{tabular}
\end{table}

%
%__________________________________________________________________

\section{Stellar modelling}

We consider both stars as isolated stars since the system is detached. Unless it is specified, the temporal evolution of the internal structure of both stars is carried out at constant mass; hence the initial mass is set equal to the observed one. Diffusion is negligible over the time duration of a PMS phase, hence we assume that the observed surface abundances correspond to the initial chemical composition. We stop the evolution when the calculated effective temperatures and luminosities correspond to the observed ones within their error bars. We stress that in the case of a binary system, one additional constraint exists: the age of both components must be the same.

Our evolutionary models are computed with the CESAM code \citep{morel97}; they are standard in the sense that no effect of magnetic field is included. Effects of rotation and diffusion are not included as the time scale of these phenomena are very long compared with the time spent by a star on its PMS (20Myr for a 2 $M_{\odot}$).

We used the OPAL equation of state \citep{rogers96} and OPAL opacities \citep{iglesias96} complemented by \citet{alexander94} opacities at low temperatures. 

During its PMS phase, a star can possess either a convective envelope or a convective core or both. The temperature gradient in convection zones is computed using the classical mixing-length theory. The mixing length is defined as $l=\alpha H_{\rm P}$, $\alpha$ being the mixing length parameter and $H_{\rm P}$ the local pressure scale height, $H_{\rm P}=-\frac{dr}{d\ln P}$. We set the $\alpha$ value equal to 1.62 as obtained for a calibrated solar model calculated with the same input physics. We define the overshooting $d_{ov}$ as the length of penetration of the convective eddies into the adjacent radiative zone: $d_{ov} = \alpha_{ov} H_{\rm P}$ with  $\alpha_{ov}$ the overshooting parameter. Unless otherwise specified, our models are computed without overshooting, i.e. $d_{ov}=0$.

The species entering the nuclear network are: $^1$H, $^3$He, $^4$He, $^{12}$C, $^{13}$C, $^{14}$N, $^{15}$N, $^{16}$O, $^{17}$O. We considered $^2$H, $^7$Li and $^7$Be in equilibrium and only the most important reactions of PP+CNO cycles are taken into account. The nuclear reaction rates are taken from  NACRE compilation (Angulo et al. 1999).

The metallicity defined as $\rm[Fe/H]=log(\frac{Z/X}{(Z/X)_{\odot}})$ where  $Z$, $X$ are the metal and hydrogen mass fractions respectively, is an observational constraint. Its initial value is taken  equal to the observed one (Table \ref{mrt}) and the abundance ratios are the solar ones 
\citep{grevesse93}. We take the initial helium mass fraction equal to the value from the solar calibration: $Y_0=Y_{\odot}=0.267$, and the hydrogen mass fraction is given by $X_0=\frac{1-Y_0}{1+(Z/X)_0}$. The atmosphere is calculated assuming the Eddington grey law.

Each evolution is initialized with a homogeneous, fully convective model in quasi-static contraction and we define the age of a star as the time elapsed since initialisation. Although our initial models are not realistic, recent studies have compared the evolutionary tracks during the PMS phase, considering two different initial conditions: the first model being calculated from the initial cloud contraction and the second one being calculated only from the Hayashi track (Cariulo et al. 2004, F. Palla, private communication). No fundamental differences between the two models were found and ages calculated in both cases are almost identical. So the initial conditions for the formation of the PMS star do not interfere in our study.

These stellar models will be considered from now on as our standard models.

\begin{figure}
\centering
\resizebox{\hsize}{!}{\includegraphics{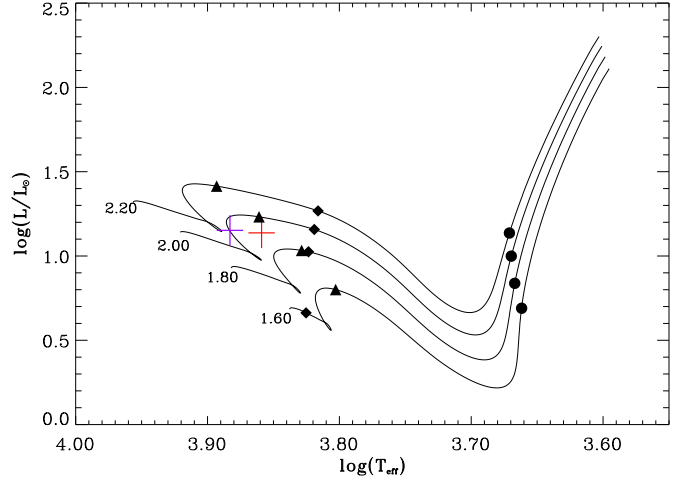}}
\caption{Evolutionary tracks  for 1.6-2.2 $M_\odot$ standard models in a HR diagram. 
Crosses represent observational error bars in effective temperature and luminosity  (Table \ref{mrt}) for  the primary (P) component (left) and for the secondary (S) component (right) of RS Cha. Dots indicate the onset of the radiative core, diamonds signal the convective envelope disappearance and triangles mark the convective core apparition.}
\label{hr}
\end{figure}

Fig. \ref{hr} shows PMS evolutionary tracks from the Hayashi line to the ZAMS for masses in the range of interest for RS Cha. We have indicated by diamonds the locus where the star becomes totally radiative and by triangles the locus where a convective core appears in the center of the star. At the stage where RS Cha is encountered, a star, evolving from the Hayashi line towards the ZAMS, has developped a fully radiative envelope and a small convective core. The earliest nuclear reactions of the CNO cycle and the p-p chain begin. The energy released by the $\rm{C}^{12}(p,\gamma)\rm{N}^{13}(\beta^+,\nu)\rm{C}^{13}(p,\gamma)\rm{N}^{14}$ chain overheats the internal regions of the star. The energy excess builds an excess pressure gradient which slows down the contraction rate in the center. At the same time, the overheating leads to the apparition of a convective core. The central regions begin to expand. The luminosity then  decreases because not only do these expanding internal shells not contribute any more to the energy flux, but they blanket the nuclear source, by absorbing a large fraction of the nuclear energy produced (see detailed description of the PMS evolution in \citet{iben65}). For later discussions, it is important to note that the primary, as it is slightly more massive, is in a more advanced stage of evolution than the secondary one. As a result, its luminosity has already decreased unlike that of the secondary one (Fig. \ref{box}).

\section{Observational contraints in a HR diagram}

\begin{figure}
\centering
\resizebox{\hsize}{!}{\includegraphics{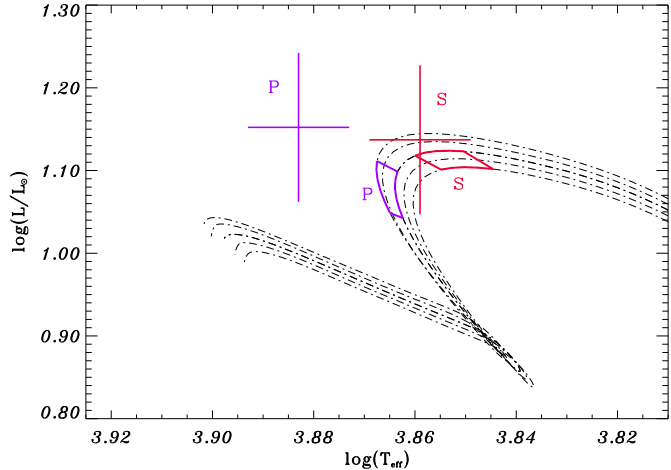}}
\caption{Evolutionary tracks of standard models (dash-dotted lines) and error boxes in masses and radii for the primary (P) and the secondary (S) components. Crosses represent observational error bars in effective temperature and luminosity (Table \ref{mrt}) for  the primary (P) component (left) and for the  secondary (S) component (right) of RS Cha. Lowest to upper tracks are for 1.86, 1.87, 1.88, 1.89 and 1.90 $M_{\odot}$. }
\label{box}
\end{figure}

Table \ref{mrt} details the physical parameters of both components of RS Cha derived from observations. Effective temperatures and luminosities as well as masses and radii will then be compared to the  corresponding calculated parameters obtained using standard models. We have calculated evolutionary tracks for the observed masses of both components as well as for their extreme values given by  the error bars. Fig \ref{box} shows the end of the PMS phase of these tracks for 1.86, 1.87, 1.88, 1.89 and 1.90 $M_{\odot}$ masses. 
 
For a given  mass $M_{obs}$, a stellar model is evolved till its radius matches the observed radius $R_{obs}$. This provides the theoretical luminosity and effective temperature $(L,T_{eff})_{calc}$ of the model. The same is done for the extreme values of the observed masses and radii  and this gives rise to error boxes which are superimposed onto the tracks in Fig.\ref{box} in full lines. In the same graph, crosses represent the error bars in luminosity and effective temperature for the primary (in left) and secondary (in right) components derived from observations $(L,T_{eff})_{obs}$.

When both crosses are located inside their respective error boxes, the corresponding stellar models  reproduce the observations. As it can see in Fig. \ref{box}, this does not occur when standard models, built as detailed in the previous section, are used. Moreover, the calculated ratio of the primary to the secondary luminosities, $(L_P/L_S)_{calc}$, is lower than one, contrary to the observations, $(L_P/L_S)_{obs}=1.05$ (Table \ref{mrt}). Our standard models indeed indicate that the luminosity of the P component has begun to decrease, unlike the luminosity of the S component. This implies that the P component has started nuclear burning earlier than the S component (see Sec. 2).

In order for the numerical models and the observations to agree, in particular to invert the $(L_P/L_S)_{calc}$ ratio,  the error box  of the primary ought to be shifted towards greater luminosities with respect to the secondary box. According to the evolutionary phase of the primary as described in the previous section, this means that it is necessary to delay the onset of the CNO cycle for the primary.

The way this can be achieved is carried out in the next section where we investigate the sensitivity of the stellar models in this evolutionary phase to various  parameters and physical inputs.

\section{Sensitivity to physical inputs and parameters}

\subsection {Unchanged error boxes in a HR diagram}

Some physical inputs and parameters of stellar model shave no effect on the location of the error boxes of RS Cha in a HR diagram:

\subsubsection {Mixing length and overshoot parameters: }

An increase in the mixing length parameter $\alpha$ moves a PMS evolutionary track towards larger effective temperature and towards slightly larger luminosities, when the star possesses a convective envelope (see \citet{bohm92}). But, as seen in Fig.\ref{hr}, the RS Cha  stars have no longer a convective envelope and changing $\alpha$ therefore has no effect on the error boxes.

Besides, the convective core just appearing in both components is too small (less than one tenth of stellar radius) to involve a modification of the tracks when changing the overshoot parameter $\alpha_{ov}$ from 0 to 0.2.

Note also that switching from a model atmosphere built assuming an Eddington grey law 
or based on Kurucz  model atmospheres \citep{kurucz79,kurucz92,kurucz93} has no effect as the envelopes of both RS Cha components are fully radiative.

\subsubsection {Equation of state:} 

Coulomb effects and departure from a perfect gas are very small for a 2$M_{\odot}$ star. Hence no variation of the tracks, whatever the evolutionary stage, is observed when using either the equation of state EFF \citep{eggleton73} or the equation of state OPAL \citep{rogers96}.
 
\subsubsection {Burning of light species:} 

The light species $^2$H, $^7$Li and $^7$Be burn at the beginning of the PMS phase during the first million years of the star, during which the gravitational energy is dominant. The energy released by the burning of light species is too small to perturb the subsequent evolutionary tracks.

\subsubsection{Mass loss:}

PMS stars of intermediate mass are known to possess stellar winds with mass loss rate in the range $10^{-8}-10^{-7}M_{\odot}.yr^{-1}$ (e.g. \citet{bouret98}). They can also accrete mass from a surrounding disk with mass accretion rates similar to those of the wind.

We have computed PMS evolutionary tracks assuming mass loss or accretion with a rate of $10^{-8}M_{\odot}.yr^{-1}$. Initial masses have been chosen such that masses of the models at the age of RS Cha ($t_{RS Cha}$) are equal to the observed masses of RS Cha. As a consequence, although the evolutionary tracks differ before and after the age of RS Cha depending on the assumed mass loss or accretion, the error boxes of RS Cha are unaffected.

\subsection {Unchanged luminosity ratio}

We now discuss the physical parameters that affect the position of the error boxes in the HR diagramme but do not allow to solve the problem of the luminosity ratio. From now on and for the sake of visibility, we have removed the evolutionary tracks from the figures. Only error boxes obtained from two different models along with crosses are kept into the HR diagram. Boxes represented by dashed lines are calculated using standard models as defined in Sect.2.1.

\subsubsection{Opacity}

\begin{figure}
\centering
\resizebox{\hsize}{!}{\includegraphics{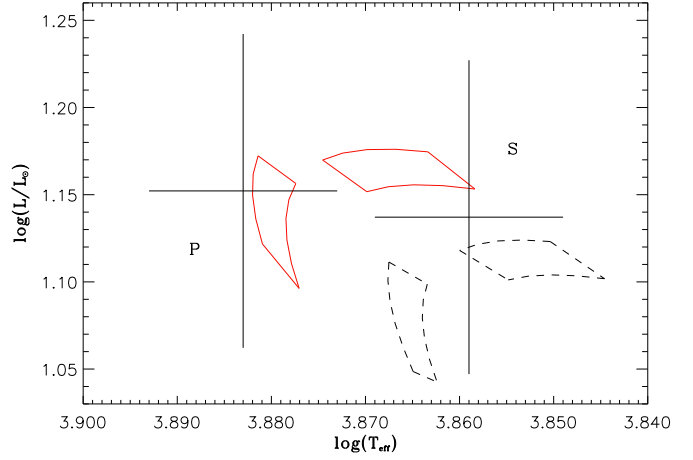}}
\caption{Error boxes calculated by decreasing  OPAL opacity of 10\% (full line) and standard OPAL opacity (dashed lines). Crosses represent observational  error bars in effective temperature and luminosity (Table \ref{mrt}) for  the primary (P) component (left) and for the  secondary (S) component (right) of RS Cha.}
\label{opacity}
\end{figure}

We change the opacity ($\kappa$) in the whole star by diminishing it by 10\%. Fig. \ref{opacity} shows that the error boxes are shifted towards larger luminosities and temperatures. A decrease in opacity makes the energy release easier and in order to maintain the pressure balance models become more centrally concentrated and hotter which leads to an increase in luminosity and effective temperature \citep{eddington26}. However both boxes move together towards larger temperatures and luminosities leaving the ratio $L_P/L_S$ unchanged.

\subsubsection{Helium mass fraction and metallicity}

\begin{figure}
\centering
\resizebox{\hsize}{!}{\includegraphics{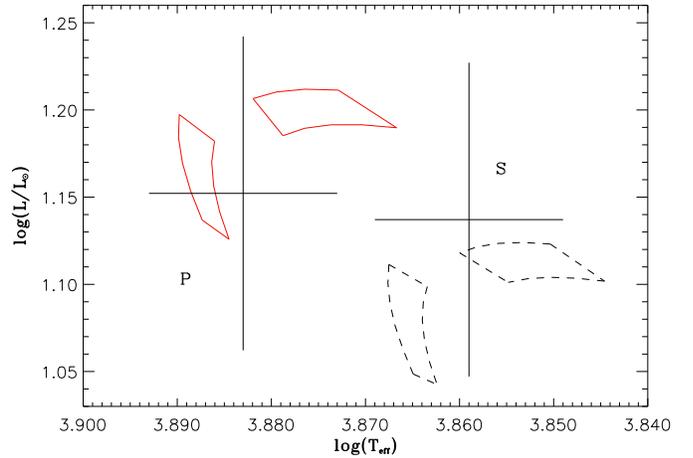}}
\caption{Error boxes calculated by increasing the helium mass fraction while keeping $Z/X$ constant: $Y=0.267$ (dashed line) and $Y=0.300$ (full line).}
\label{Y}
\end{figure}

In paper I, we have determined the metallicity $\rm[Fe/H]$ of RS Cha (Table \ref{mrt}), assuming that both stars have the same chemical composition and that the metal abundance ratios are the solar ones. The knowledge of $\rm[Fe/H]$ gives us a constraint on $\frac{Z}{X}$ for RS Cha. In order to know fully the composition of the star, we have to assume the helium mass fraction $Y$ which is unknown for this system.

We have varied $Y$ between the solar value: $Y=0.267$ and $Y=0.300$ which is a reasonable range 
compared to the observed values in various stars and clusters. Fig. \ref{Y} shows that the luminosity and the effective temperature increase with increasing $Y$. An increase of $Y$ leads to an increase of the mean molecular weight $\mu$. In order to maintain the pressure balance the star is more condensed and hotter which results in an increase in luminosity and effective temperature \citep{iben65}.

We have also varied the metallicity of the system $\rm[Fe/H]$ within the  error bars. Since heavy elements contribute only for a small mass fraction, an increase in heavy element abundances does not influence the mean molecular weight $\mu$ very much. However an increase in metallicity increases the global opacity ($\kappa$). As seen in the previous section, an opacity increase in the whole star leads to  a decrease in effective temperature and luminosity in both stars.

As both boxes  move together in the same direction, we cannot reverse the luminosity ratio by varying the helium mass fraction or the metallicity.

\subsection{Inverting the luminosity ratio}

\begin{figure}
\centering
\resizebox{\hsize}{!}{\includegraphics{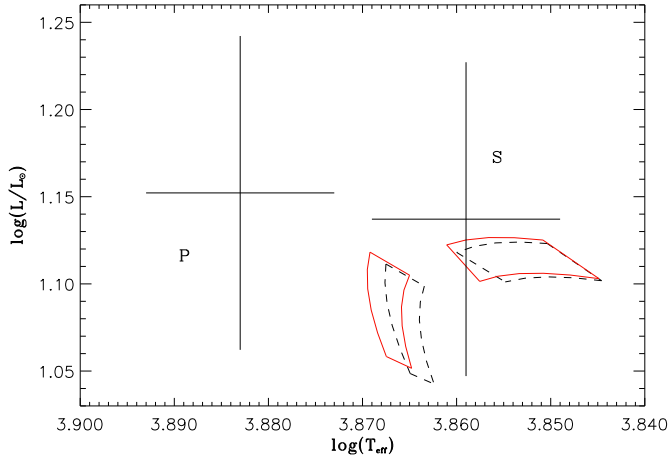}}
\caption{Error boxes calculated by decreasing the rate of the $^{12}$C(p,$\gamma$)$^{13}$N(e$^+$,$\nu$)$^{13}$C reaction of 10\% (full line) from the NACRE rate (dashed line)}
\label{nuc}
\end{figure}

So far no parameter, which changes directly or indirectly the luminosity transfer into the star, 
succeeded in inverting the calculated luminosity ratio of the  first component to the second one. We now focus on the energy generation. 

The evolutionary stage of RS Cha corresponds to the begining of the CNO cycle with the burning of carbon and nitrogen. The $^{12}$C$\rightarrow ^{14}$N chain overheats the internal regions 
of the star leading to the apparition of a convective core responsible for the luminosity decrease. The onset of the convective core for the primary must be delayed in order to invert the luminosity ratio; this means a decrease of  the efficiency of overheating of the internal regions. This efficiency depends on both nuclear reaction rates and the abundances of the species involved in the CNO cycle. 

\subsubsection{Nuclear reaction rates}

We first decrease the nuclear reaction rates by 10\%, which corresponds to their estimated  accuracy \citep{angulo99}. Fig. \ref{nuc} shows the shift of the boxes resulting from the decrease of the rate of the $^{12}$C(p,$\gamma$)$^{13}$N(e$^+$,$\nu$)$^{13}$C reaction. The luminosity of the primary comes closer to that of the secondary, but not sufficiently to invert the luminosity ratio. The decrease of the rates of the other reactions of the chain $^{13}$C$(p,\gamma)^{14}$N shows a similar, although smaller, shift.

\subsubsection{Abundances of Carbon, Nitrogen and Oxygen}

\begin{figure}
\centering
\resizebox{\hsize}{!}{\includegraphics{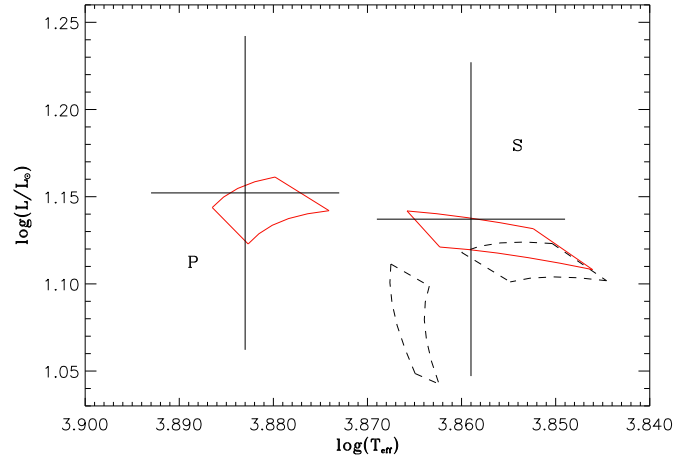}}
\caption{Error boxes calculated by decreasing the abundance of carbon: $\Delta n_C = -0.7$ dex (full line), keeping the metallicity constant.}
\label{abrat}
\end{figure}

We then decrease the abundance of carbon, nitrogen and oxygen, changing only the abundance ratios of heavy elements while keeping the metallicity unchanged. At the evolutionary stage of RS Cha the variation of the abundance of $^{17}$O is very weak, compared to those of C and N, and the abundance of $^{16}$O does not vary. The reason is the low rate of the reactions $^{16}$O(p,$\gamma$)$^{17}$F($\beta^+\nu$)$^{17}$O and $^{17}$O(p,$\alpha$)$^{14}$N (NACRE, Angulo et al. 1999). Hence we do not consider the oxygen abundance any further.
  
Hereafter we call $n_i$ the logarithm of the abundance by number $N_i$ of a chemical element $i$ in a scale where the abundance of hydrogen by number is $N_H=10^{12}$, that is
\begin{equation}
	n_i=\log\frac{N_i}{N_H}+12
\end{equation}

\begin{figure*}
\begin{minipage}[t]{8.5cm}
\centering
\includegraphics[width=8.5cm]{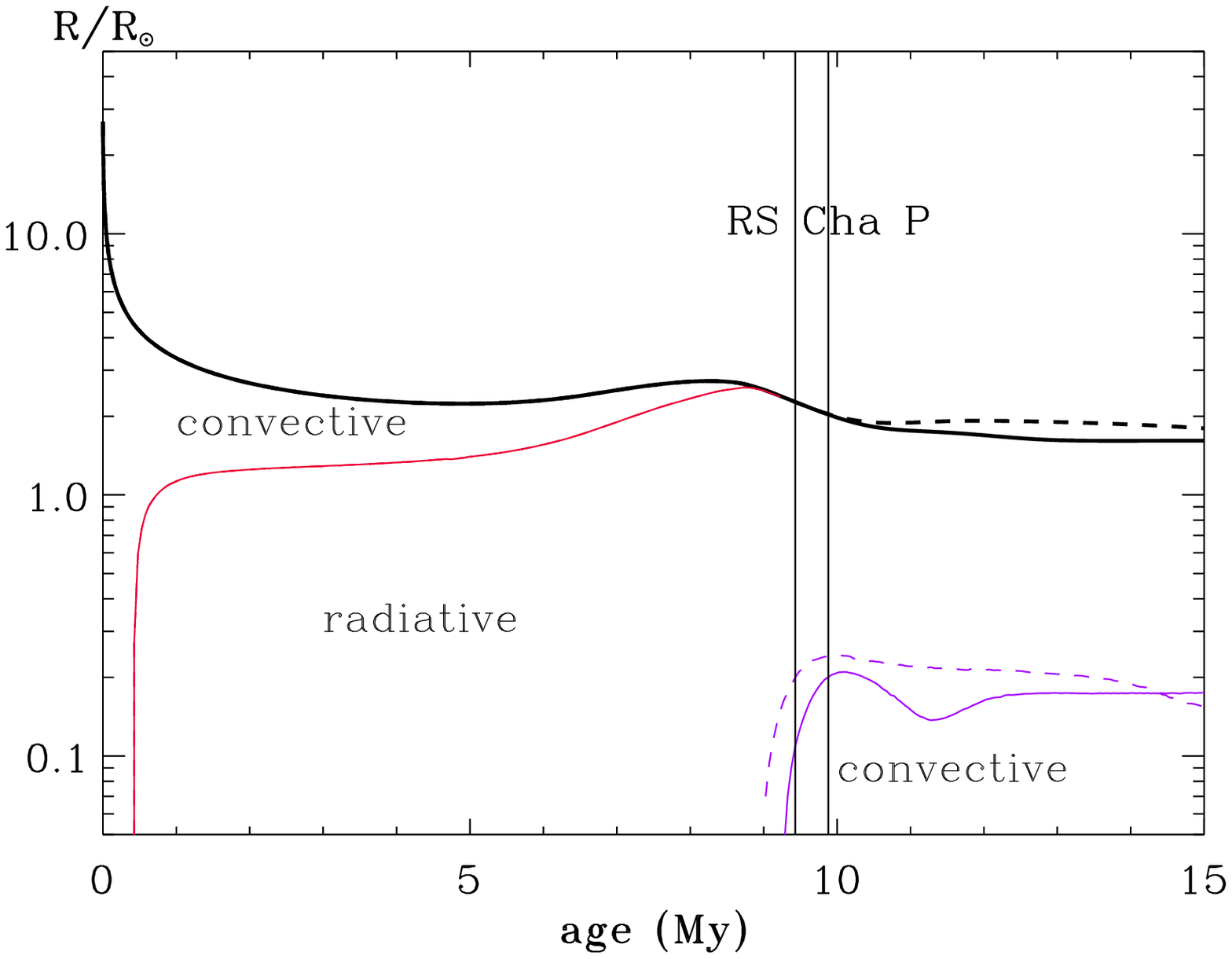}
\end{minipage}\hfill
\begin{minipage}[t]{8.5cm}
\centering
\includegraphics[width=8.5cm]{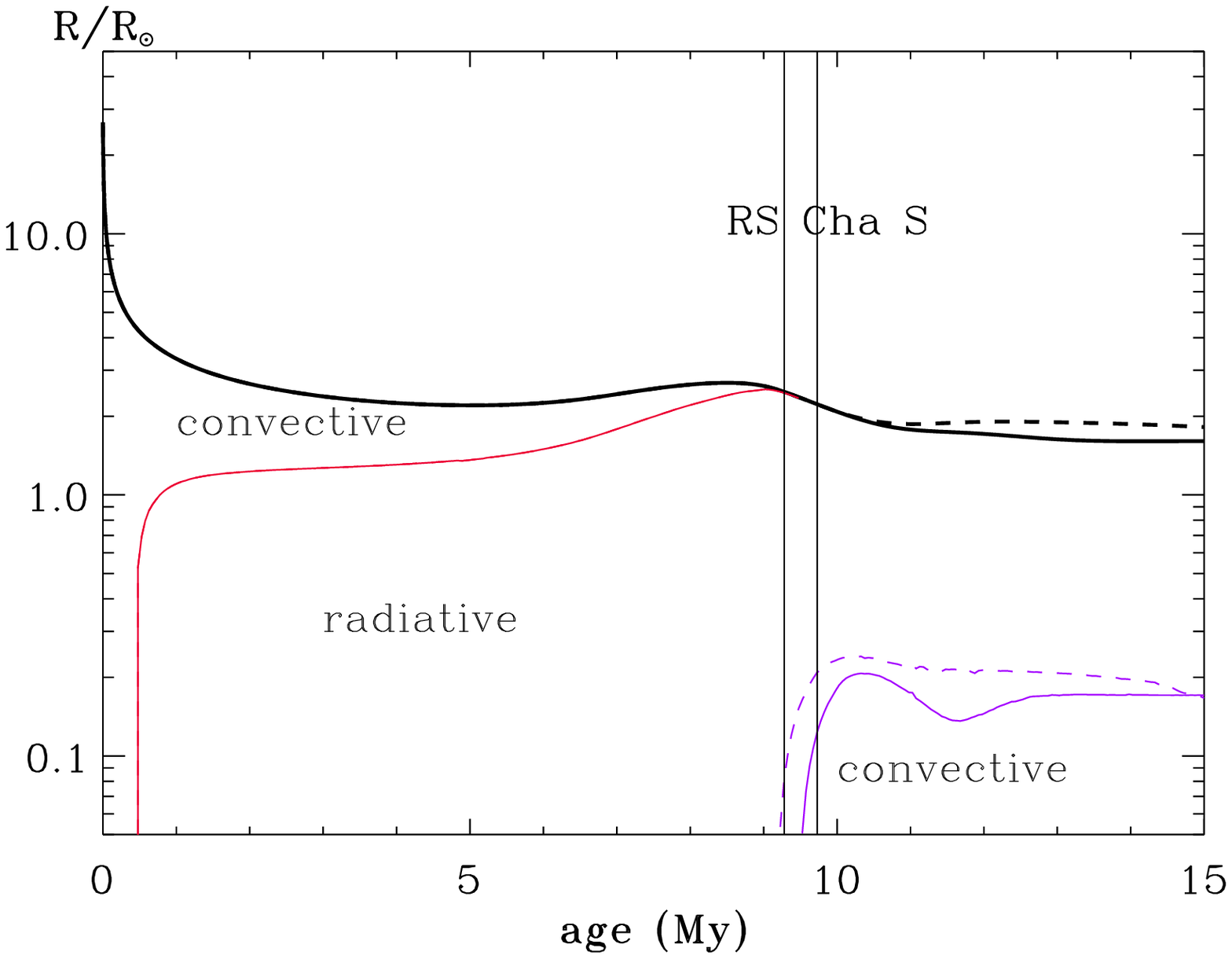}
\end{minipage}
\caption{Radius of the primary (left) and secondary (right) star plotted as a function of the age of the standard model 1 in thick-dashed-line, and model 2 built with a decrease of 0.7 dex of carbon in thick-full-line. Before $\sim$10 Myr both radii are superimposed. The thin lines represent the limit between radiative and convective zones in model 1 (thin-dashed-line) and in model 2 (thin-full-line). The two vertical full lines represent the extrema of the age of the primary and secondary components of RS Cha (RS Cha P and RS Cha S) calculated from the error boxes P and S of Fig. \ref{abrat}. We see that the convective core in the secondary component is less developped than that of the primary one}
\label{conv}
\end{figure*}

Fig. \ref{abrat} shows the shift of the error boxes resulting from a decrease of the carbon abundance $n_C$ by 0.7 dex. As we remove some carbon, the reactions $^{12}$C$\rightarrow ^{13}$C and $^{13}$C$\rightarrow ^{14}$N become less efficient. The nuclear energy released $\epsilon$ is proportional to $L/m$, $L$ and $m$ being the local luminosity and mass;  the radiative gradient $\nabla_{rad}$ is proportional to $\frac{L}{m}\frac{\kappa P}{T^4}$, where $\kappa$, $P$ and $T$ are the local opacity, pressure and temperature (e.g. \citep{kippenhahn90}). We find that decreasing the abundance of carbon leads to a negligible increase of $\kappa$, $P$ and $T$ in the center of the star, compared to the large decrease of the $L/m$ ratio. Therefore, by reducing the carbon abundance, the drop of the produced nuclear energy  leads to  a dominant decrease of the $L/m$ ratio, and therefore to a decrease of the radiative gradient $\nabla_{rad}$ which delays the onset of the convective core. Fig. \ref{conv} shows the variation with time of the radius of each star in model 1 (normal carbon abundance) and model 2 ($n_c$ depleted by 0.7 dex). While at ages lower than $\sim$9 Myr, the radii of the models as well as the limit between the radiative core and the convective envelope are the same, at ages higher than $\sim$9 Myr -after the onset of the CNO cycle- the convective core appears later and remains smaller in model 2.

As a result, the expansion of the internal shells is delayed, as well as the decrease in total luminosity (see Sec. 2). The evolutionary track is extended towards larger luminosities and temperatures and we therefore obtain a luminosity ratio ($\frac{L_P}{L_S}$) greater than one, as observed. We notice the same phenomena with the decrease of the nitrogen abundance, which makes the reactions $^{14}$N$(p,\gamma)^{15}$O$(\beta ^+,\nu)^{15}$N and $^{15}$N$(p,\gamma)^{16}$O less efficient.

\section{Calibration of RS Cha}

We have finally found that the luminosity ratio $\frac{L_P}{L_S}$ is sensitive to the abundances of carbon and nitrogen. We now search a model matching properly  the observations. This model is not very different from our standard model. We must change only the abundance ratios of heavy species and the helium mass fraction. Our final models are obtained by decreasing the carbon and nitrogen abundances by 0.6 and 0.5 dex respectively. The decrease of 0.6 dex and 0.5 dex of the number of atoms of carbon and nitrogen gives new abundances equal to 8.25 and 7.77, respectively (the solar values are 8.55 and 7.97). \citet{daflon04} measured the abundances of various species in 69 young OB stars in 25 clusters. They found abundances comprised between 8.21 and 8.49 for carbon and between 7.19 and 7.83 for nitrogen, showing that our values agree well with abundances observed in young stars. We should also note that our results are going in the same direction as the recent solar abundances determination of Asplund et al. (2004,2005a,b). They found a decrease of 0.16 dex of the carbon abundance and of 0.11 dex of the nitrogen abundance and of 0.21 dex of the oxygen abundance, compared to the abundances of \citet{grevesse93} used in the CESAM code.

\begin{figure*}
\centering
\includegraphics[width=17cm]{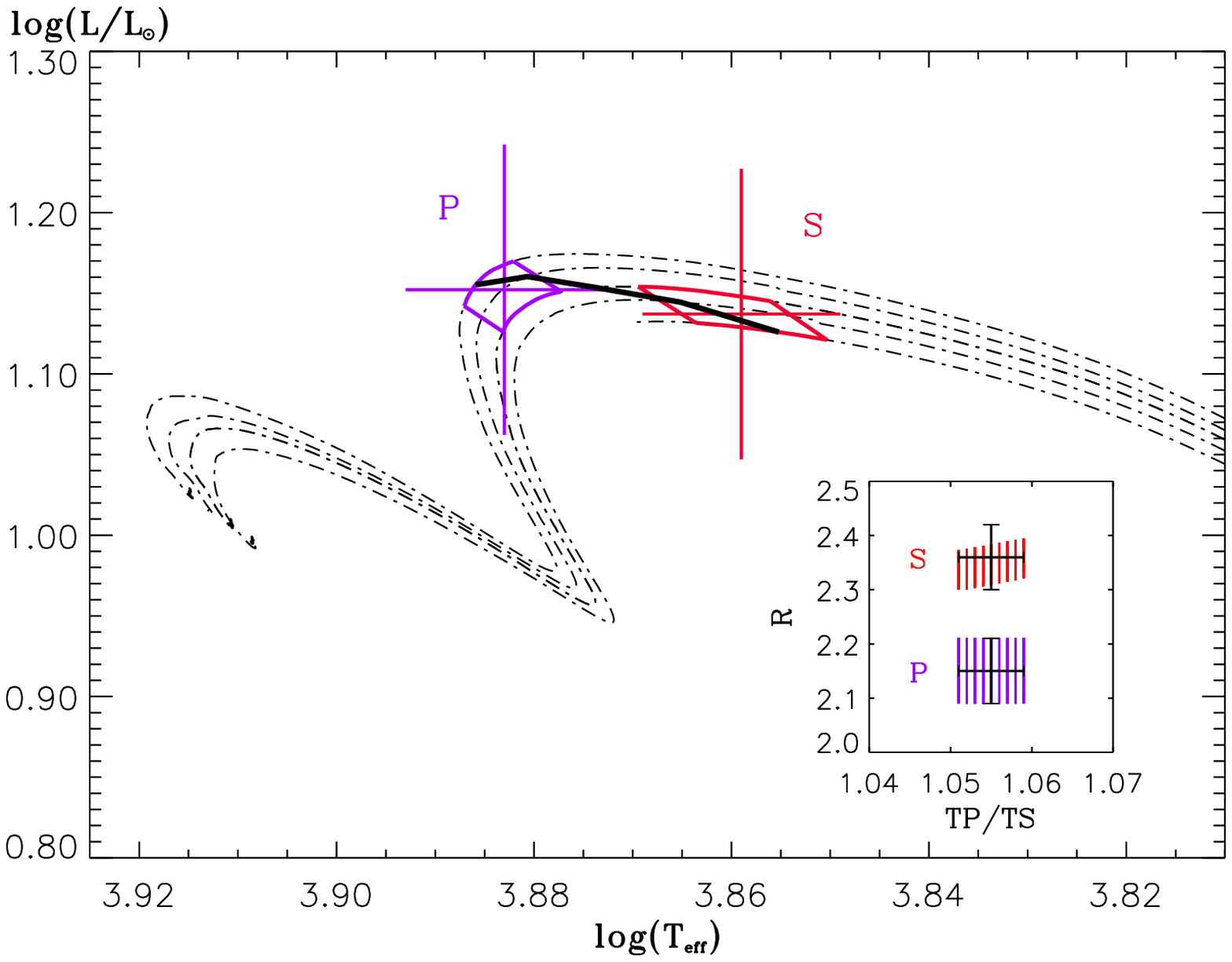}
\caption{Evolutionary tracks (dash-dotted lines) and error boxes (full lines) calculated by decreasing the abundances of carbon and nitrogen: $\Delta n_C = -0.6$ dex and $\Delta n_N = -0.5$ dex. The thick line is the $9.50~ Myr$ isochrone. Crosses represent observational  error bars in effective temperature and luminosity (Table 1) for the primary (P) component (left) and for the  secondary (S) component (right) of RS Cha. The lower right panel shows the constraints on the radii of both stars when the photospheric measure of the effective temperature ratio of \citet{clausen80} is taken into account. Crosses represent the error bars in radii and temperature ratio for both stars. Each line in the hatched area represents the allowed values of R for a given value of $T_{\rm P}/T_{\rm S}$.}
\label{abratCN}
\end{figure*}

Having the appropriate luminosity ratio between both components of the system, we need to increase the helium mass fraction in order to match the calculated parameters of RS Cha to the observed ones (Table \ref{mrt}). The error bars in masses and radius of the models are plotted in Fig. \ref{abratCN}. We find a value of $Y= 0.272$ ($\log(N_{He}/N_H)+12=10.99$), reasonable compared to the helium mass fraction observed in young stars \citep{mathys02}.

\medskip 
Finally, it remains to be checked that both stars have the same age. We find that a common interval of ages does indeed  exist inside the error boxes indicating that our final models for both components can have the same age. We illustrate this by plotting the 9.50 Myr isochrone crossing both boxes in Fig. \ref{abratCN}. This age is of the same order as previous determinations \citep{mamajek00,luhman04}. However we stress our age determination is based on an observed metallicity contrary to previous works.

Besides, in a binary system the ratio of the effective temperatures is better constrained than the absolute values of the temperatures of the individual stars. \citet{clausen80} measured very accurately the photometric value of this ratio : $T_{\rm P}/T_{\rm S}=1.055\pm0.004$. In order to take into account this additionnal constraint, we checked that our calibration model can reproduce simultaneously this ratio and the observed radii of both stars, by plotting error boxes in the radius versus $T_{\rm P}/T_{\rm S}$ diagram as follows. We first calculated the effective temperatures ranges delimited by the error boxes in masses and radii of Fig.\ref{abratCN} for the primary and secondary components on the same isochrone. Then we restricted these intervals to the primary and secondary temperatures which satisfy the observed ratio of $T_{\rm P}/T_{\rm S}$ within the error bars. Finally we calculated by linear interpolation the corresponding radii ranges. We obtain two boxes as shown in the lower right panel of the Fig. \ref{abratCN}. We find that our calibration models reproduce the observed radii of both stars as well as the observed effective temperature ratio of the system.
\\[0.5 cm]
\indent We have seen that the small area of the PMS tracks where RS Cha is observed is only weakly sensitive to various physical inputs and corresponding stellar models are then only weakly constrained. Fortunately more severe constraints can be obtained  by a seismic sounding of the stellar interior, provided the stars oscillate.

%
%______________________________________________________________

\section{RS Cha: a binary system of pulsating PMS stars}

\subsection{Observations of delta-scuti type pulsations}

Many authors \citep{andersen75,mcinally77,palla01} discussed the possibility of the presence of oscillations in one of the components of RS Cha but reported no direct observations of these oscillations. Andersen (1975) mentioned hints of variability in the residuals from the observed radial velocity curve of the primary but could not be more conclusive with the data at his disposal. Marconi \& Palla (1998) calculated the theoretical location of the instability strip of PMS stars in the HR diagram for radial modes. Palla \& Stahler (2001) located RS Cha in the HR diagram and found that the secondary component lies inside the instability strip and ought to  be a PMS $\delta$ Scuti star. The primary component was found  outside but near the left of the instability strip for fundamental radial mode which cannot exclude that the star is pulsating  in  radial  overtones and/or  nonradial modes.

Finally, using our data, we have recently shown, that {\bf both} components of RS Cha are probably pulsating (paperI). We have pointed out temporal variations in the residuals from radial velocity curves of both stars.  These variations may be periodic with  periods around one hour. We therefore ascribed them to $\delta$ Scuti type oscillations. 

Unfortunately our data were not accurate enough and did not cover sufficient time to determine precisely the periods. However we can wonder whether the period around one hour found with our data belongs to the period range of the excited modes that can be expected theoretically.

\subsection{Theoretical oscillation modes and periods}

\begin{figure*}[t!]
\begin{minipage}{8.5cm}
\centering
\includegraphics[width=8.5cm]{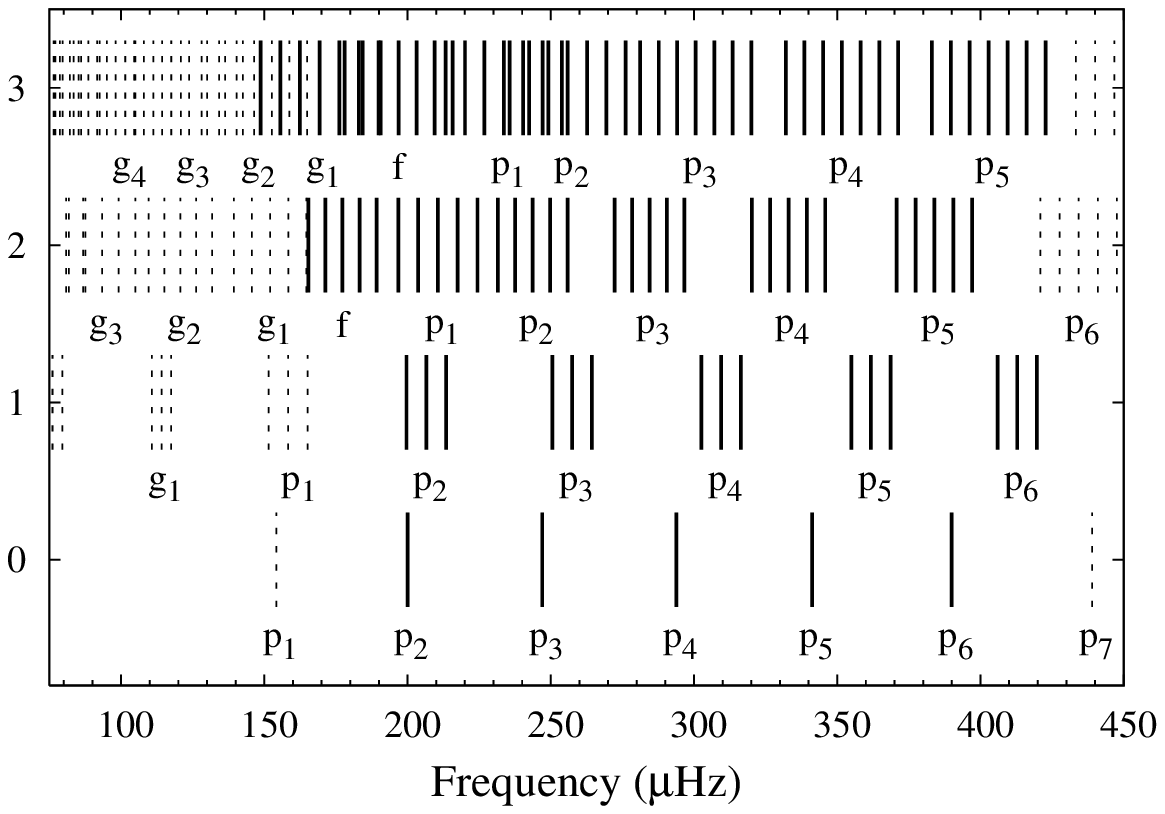}
\end{minipage}\hfill
\begin{minipage}{8.5cm}
\centering
\includegraphics[width=8.5cm]{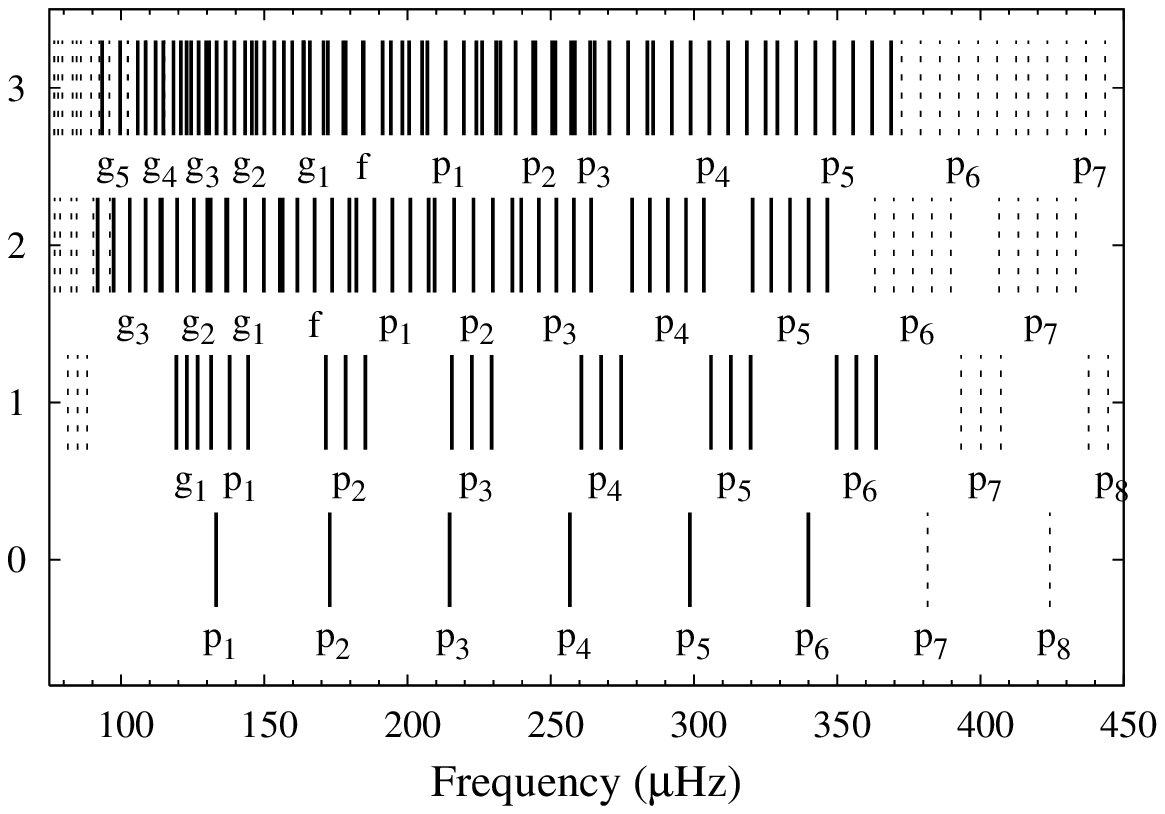}
\end{minipage}
\caption{\footnotesize
Theoretical frequencies of pulsation modes obtained with non-adiabatic calculations, for the primary (left) and the secondary (right) stars. Each bar represents one mode  as a function of its frequency (in $\mu Hz$) and its degree $\ell$ (vertical direction). Dashed bars represent  stable modes and  solid bars unstable modes. Label $p$ is for $p$ mode and label $g$ for $g$ mode. Numbering of the label $p$ and $g$ referes to the radial order of the mode}
\label{mode1}
\end{figure*}

Theoretical periods of pulsation modes of both components of RS Cha were calculated using Dupret's code MAD \citep{dupret05}.  This nonadiabatic code includes a time-dependent interaction between pulsation and turbulent convection \citep{grigahcene05}. The blue and red edges of the instability strip are accurately determined with this code for radial and non-radial modes of MS and PMS stars \citep{dupret05,grigahcene06}.

The stellar parameters of the selected models for RS Cha components  are those resulting from our calibration (see Sec. 5).

In paper I, it was shown that the system is synchronized and the orbital period of the system has been determined accurately, we therefore know the rotational period of both stars: $P_{rot}= 1.67$~d. For sake of simplicity we assume a uniform rotation  with the surface  angular velocity $\Omega= 2 \pi /P_{rot}$. This enables us to compute the rotational splitting  for each mode, according to the first order perturbation theory \citep{ledoux58}
\begin{equation}
	\nu_{nlm} = m \Omega (1-C_{n\ell})
\end{equation}
in the observer reference frame. $C_{n,\ell}$ is the Ledoux constant. These $m\not=0$ components of the multiplets are included in the frequency set and they are assumed stable or unstable according to the same status of the associated $m=0$ mode.

Fig. \ref{mode1} shows the frequency interval over which modes are unstable and therefore expected to be detected. Comparison between the plots of each component illustrates a well known result for $\delta$ Scuti like pulsations: unstable modes have higher frequencies and radial order towards the blue side of the instability strip (here the primary component).

We find that  the periods of excited modes are around one hour for both stars (those of the secondary are a little longer than in the primary). On one hand, this supports the conclusion that the observed period corresponds to a low order pressure mode of delta-Scuti type and on the other hand this shows our models are not too far from reality.
\medskip 

In Sect. 4.1.1., we have seen that including convective core overshoot ot not does not change  the location of the RS Cha models in the HR diagram indicating that the structures of the models are not too much affected when considering these photometric observables. We now consider the effet of including overshoot on the frequencies of each component of the system separately  as an indication of the efficiency of the seismic diagnostic ability of RS Cha system.   

\begin{figure}[t!]
\centering
\includegraphics[width=8cm]{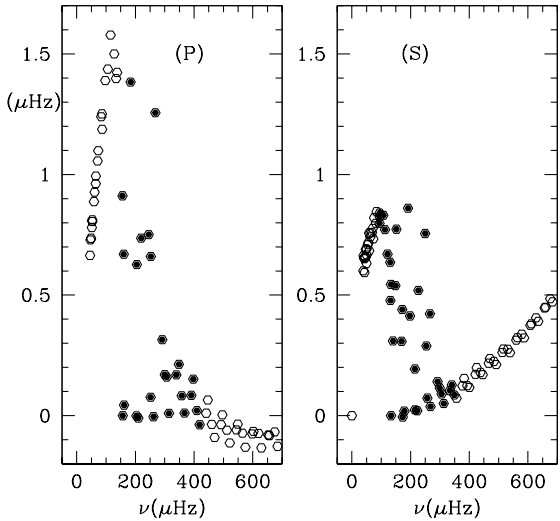}
\caption{\footnotesize Frequency differences as a function of the frequency  for each   oscillation mode $(n,\ell)$ between frequencies of one stellar model with overshoot and one without  overshoot for  the primary RS Cha component on the right and the secondary RS Cha component on the left.  Black (resp.open) dots represent unstable (resp. stable) modes}
\label{diffnu}
\end{figure}

For each component of RS Cha, we  compare the frequencies for each mode $n\ell$ computed from two models which differ by the value of the  overshoot parameter: one model has no overshoot $d_{ov}=0$ and the other one is computed assuming $d_{ov}=0.2 H_p$. These models selected for comparison have the same mass and are located quite close to each other in the HR diagram. Fig.\ref{diffnu} shows frequency differences
\begin{equation}\label{eqdiffnu}
	\nu_{n,\ell}^{(ov)}-\nu_{n,\ell}^{(noov)} {\nu_{1,0}^{(ov)}\over \nu_{1,0}^{(noov)}}
\end{equation}
as a function of $\nu_{n,\ell}^{(noov)}$. Frequencies $\nu_{n,\ell}^{(ov)}$ refer to frequencies for each oscillation mode $(n,\ell)$ computed from the stellar model with convective core overshoot and $\nu_{n,\ell}^{(noov)}$ those for the stellar model computed assuming no convective core overshoot. The ratio ${\nu_{1,0}^{(ov)}\over \nu_{1,0}^{(noov)}}$ is a scaling factor which corrects for the fact that the  models with and without overshoot do not have the same radius.

Differences at high frequency for the secondary arise because the selected models with and without overshoot for this component have not quite the same radius and this is not completely corrected by the scaling $\nu_{1,0}^{(ov)} /\nu_{1,0}^{(noov)}$. On the other hand at low frequency, differences are large for the primary as the models which have been chosen for comparison have the same radius but not quite the same evolutionary status. Fig.\ref{diffnu} indicates that large frequency differences arise for modes which are in the unstable range for both stars and if detected could be quite discriminant.

%
%______________________________________________________________

\section{Conclusions}

In this paper, we carried out  a stellar modelling of a binary system whose physical parameters - masses, radii, luminosities and effective temperatures are well constrained by observations. The RS Cha system is composed  of two   similar intermediate mass stars in a PMS stage of evolution quite close to the ZAMS at the onset of CNO burning. Assuming that both stars have the same chemical composition, a first modelling using the spectroscopically determined  iron abundance for this system and  assuming solar abundance ratios could not satisfy the observational constraints simultaneously for both stars.

In order to obtain a consistent modelling of both stars, we investigate the sensitivity of the models in the evolutionary stage of the RS Cha system - at the end of the PMS phase - to various uncertainties in the modelling, including uncertainties in the input physics and initial abundances. We found that the stellar structure is only weakly sensitive to uncertainties in the stellar physical description.

The calibrated models i.e. the final ones which satisfy all the observational constraints 
were obtained only by changing the abundance ratios of carbon and nitrogen which then are in agreement with known abundances of other young stars. Once the chemical composition is set as well as the other stellar parameters intrinsic to RS Cha stars, the error bars on the effective temperatures and luminosities, although already quite small, remain still too large to put  the standard  physical input of our stellar models at default. These models also satisfy the constraints on the effective temperature ratio.

We find that there is no need to include rotation in our models as they are able to reproduce the observed properties of RS Cha within their error bars. We note that similar conclusions are drawn by \citet{claret06} in their study of EK Cep binary system. Furthermore we have checked that the effects of rotation on the global parameters in the HR diagram ($L,T_{\rm eff}$) estimated from the grid of rotating atmospheres of \citet{hernandez99} for the $v\sin i$ values of the components (Table 1) are negligible with respect to the error bars.

In order to obtain useful constraints on the physical description in this evolutionary phase, 
we therefore need to find other properties of stars than the photometric and spectroscopic ones, that-is seismic properties. The binary system RS Cha is a good example as we have recently proved that both components are pulsating. The detected variations indicate periodicities around 1 hour which we find to fall within the range of theoretically opacity driven  modes with radial orders p$_3$-p$_4$ for the primary and  p$_4$-p$_5$ for the secondary. We find  that in this range the frequency spectra for both stars are regular  when  including modes with spherical degrees $\ell=0,2$, an advantage when dealing with mode identification  and possible seismic inferences \citep{suran01}.

For a deeper study however, more observations are needed in order to determine the seismic properties of both stars more precisely. This is not an easy task. This binary system is not resolved even with our biggest telescopes. We cannot therefore obtain separate photometric light curves. We can only observe the spectroscopic variations of the radial velocities of both stars. Getting modes and periods by this way, requires long-term multi-site monotoring (which is under preparation).

\begin{acknowledgements}
We are very grateful to Francesco Palla for fruitful discussions and for providing us with the ages of the models computed with his FRANEC code. We warmly thank the referee for its judicious remarks.
\end{acknowledgements}

\nocite{*}
\bibliographystyle{aa}
%\bibliography{rschamod}

\end{document}